\documentclass[11pt,a4paper]{article}
\usepackage{jheppub}
\usepackage{physics}
\usepackage[table]{xcolor}
\usepackage{multirow}
\makeatletter
\renewcommand*\env@matrix[1][*\c@MaxMatrixCols c]{%
  \hskip -\arraycolsep
  \let\@ifnextchar\new@ifnextchar
  \array{#1}}
\makeatother

\title{Double SU(4) model}
\author[1]{Jose A.R. Cembranos}
\author[1,2]{and Pablo Diez-Valle}

\affiliation[1]{Departamento de F\'isica Te\'orica and IPARCOS, Universidad Complutense de Madrid, E-28040 Madrid, Spain}
\affiliation[2]{Consejo Superior de Investigaciones Cient\'ificas (CSIC), E-28006 Madrid, Spain}
\emailAdd{cembra@ucm.es}
\emailAdd{pablodie@ucm.es}
\abstract{In this manuscript we study the Double SU(4) model as a grand unified theory based on the gauge group $\,SU(4)\times SU(4)\left(\times \mathcal{Z}_2\right)$. A complete set of generators is constructed according to a pattern of symmetry breaking via a Pati-Salam stage. Furthermore, the gauge boson matrices are derived and the extended Gell-Mann Okubo relation is calculated. All known elementary particles and antiparticles, besides exotic degrees of freedom, are unified in multiplets which are transformed as $(4,4)$ and $(\bar{4},\bar{4})$ under the proposed gauge group.}

\begin{document}
\maketitle
\flushbottom

\section{Introduction}
\label{Introduction}

The Standard Model of particles physics (SM), which describes the strong, weak and electromagnetic interactions, has shown to be very successful, especially after the detection of the electroweak Higgs in the Large Hadron Collider (LHC) at CERN in 2012 \cite{CMShiggs,Higgsdiscovery}. The SM is able to explain a huge number of phenomena at the subatomic level, is consistent with almost all experimental data, and furthermore, all the particles that it predicts have now been found.\\

Despite the fact that the Standard Model is ultraviolet complete, there are considerable indications that the SM is not the ultimate theory of elementary particle
interactions and the existence of physics beyond this model. One of the main absences in SM is a mechanism 
to give mass to neutrinos. The neutrino masses are known to be non-zero \cite{Neutrinooscillations1,Neutrinooscillations2,Neutrinooscillations3,Neutrinooscillations4,Gonmichele}, but they remain zero even after the electroweak symmetry is broken. Interesting proposal are based on the so called seesaw mechanism which needs to add new degrees of freedom to the SM \cite{Mohapatra}. Another choice is to add to the SM spectrum the right-handed singlet neutrinos which would allow neutrinos to get Dirac masses after the Higgs acquires its vacuum expectation value (VEV). \\    

Some of the current SM problems related with Cosmology are the lack of a Dark Matter candidate and of a mechanism for generating the baryon asymmetry of the universe. In addition, there is no reason for the observed smallness of $\theta$ parameter of QCD. Beside these conceptual troubles, the apparently complex representations in which the matter fields arrange themselves appear to be random. Furthermore, the arbitrariness in some SM aspects disfavors this model as a fundamental theory of elementary particles. For instance, the experimental fact that the electric charge is quantized requires an explanation which some GUTs models are able to provide. \\ 

We can also see its unattractive arbitrariness in the 19 SM parameters (without taking into account neutrino masses) which must be fitted to data: three gauge couplings, 13 parameters associated with the Yukawa couplings (9 charged fermions, three mixing angles and one CP phase in the CKM matrix), the Higgs mass and its quartic coupling, and $\theta_{QCD}$ \cite{Patrignani}. A satisfactory fundamental theory of Nature should be, in our view, more independent of data adjustment. \\

Since the 70's it is known that some of these limitations of the SM could be solved considering a gauge theory with invariance under a larger group $G$ than the SM gauge group $G_{SM}=SU(3)_C \times SU(2)_L \times U(1)_Y $.
Thereby the Standard Model would be the low-energy limit of this new gauge theory. It is the so called \textit{Grand Unified theories} (GUTs) \cite{Langacker,Ross}. In GUTs the strong, weak and electromagnetic interactions are supposed to be actually different manifestations of the same fundamental interaction, being only one at a high enough  energy scale, i.e the three SM gauge couplings are quantitative unified and a unique gauge coupling exists at the GUT energy scale.\\

The cornerstone of Grand Unification is to embed the Standard Model gauge group $G_{SM}$ in the larger local Lie group $G$. Thus, the additional symmetries may explain or may restrict some of the aspects that appear to be arbitrary or random in the SM. In addition, this embedding can solve some SM conceptual troubles previously shown such as the baryon asymmetry of the universe. The unified symmetry $G$ is broken, for instance by a Higgs mechanism, in one or more stages, recovering the SM invariance under $G_{SM}$ at low energy scales. The SM gauge couplings are unified in a unique parameter if $G$ is a simple Lie group. Therefore, the three forces are only one interaction at energy scales at which the $G$ symmetry is not broken. Then, the different features of the strong, weak, and electromagnetic interactions, as well as their gauge couplings, would be determined by the spontaneous symmetry breaking pattern of the model. \\     

One of the first attempts of unification of the three SM forces is the Pati-Salam model (PS), a partial unified model proposed by Jogesh C. Pati and Abdus Salam \cite{Pati-Salam,Salam-Pati}. In this model the SM gauge group is embedded into the larger local group  
$G_{PS}= SU(4)_C\times SU(2)_L \times SU(2)_R$.
Since $G_{PS}$ is not a simple Lie group, it keeps different gauge couplings. In particular, the Pati-Salam Model include two gauge couplings if a discrete parity symmetry between left-handed and right-handed particles is considered: $g_C$ corresponding to the $SU(4)_C$ coupling and $g_{W}$ corresponding to $SU(2)_{L,R}$. This model groups
baryonic quarks ($B=1$) and leptons ($L =1$) together
as members of the same fermionic multiplet of $SU(4)_C$ with a defined fermionic number ($F=B+L=1$). Thus, leptons are a fourth color (generally known as white or lilac) added to the classic three quark colors (red, blue and green). In addition, the PS includes a local symmetry $SU(2)_R$ under which right-handed particles are not singlets incorporating the right-handed neutrino (missing in the SM) in a $SU(2)_R$-doublet with the right-handed electron. In this work we shall study an extension of the PS based on the gauge group $SU(4)\times SU(4)\left(\times \mathcal{Z}_2\right)$. 

\section{Double SU(4) as Grand Unified theory}

A particular choice of the grand unified group $G$, which has to content the Standard Model group, is the direct product of identical simple Lie groups. In this manuscript we study a Grand Unified theory based on the following gauge group:

\begin{equation}
SU(4) \times SU(4) \left(\times \mathcal{Z}_2\right)
\label{doubleSU4}
\end{equation}
where $\mathcal{Z}_2$ is a discrete symmetry which interchanges the two SU(4) factors. We can include $\mathcal{Z}_2$ in order to get a total unification of all gauge coupling parameters at high energy, instead generally they are independent. \\  

This model is a generalization of the classical partial unified  theory of Pati-Salam introduced in the previous section following a similar approach to \cite{Rajpoot}. The choice of this group for our model allows to keep the Pati-Salam achievements, like electromagnetic charge quantization or the lepto-quark unification but, moreover, we unify all gauge coupling constants and we express the parity violation in the electroweak theory as a consequence of the spontaneous symmetry breaking, unifying all particles in two identical representations of (\ref{doubleSU4}).\\ 

GUTs which involve $SU(4)$ factors as fundamental unified gauge group have not been extensively studied \cite{Bernardini,Cho,Rajpoot,Rajpoot2}. This is due to the fact that a single $SU(4)$ can not contain the Standard Model ($G_{SM}$), and that a semi simple group does not unify all gauge couplings without an additional discrete symmetry. This manuscript is different to previous works in $SU(4)$ GUTs because we consider a minimal particle assignment derived from the explicit construction of the group and the extended Gell-Mann Okubo relation in the framework of the Double SU(4) GUT.\\

We have organized the manuscript as follows. First of all, in section \ref{patternsym}, we shall study the patterns of symmetry breaking that we will consider. Then, based on these breaking patterns we shall present the construction of the group (\ref{doubleSU4}) in section \ref{Groupconstruction}. In section \ref{Gaugebosons} we will study the gauge bosons of the model and in section \ref{quanumbersection} we shall derive the extended Gell-Mann Okubo relation. Lastly, we shall study the fermion content in the framework of the Double SU(4) GUT in section \ref{Fermioncontent}. The main conclusions are discussed in the last section.  

\section{Pattern of symmetry breaking}
\label{patternsym}

There are several options to break the $SU(4) \times SU(4)$ group. Some straightforward choices are, for example,
\begin{equation*}
SU(4) \times SU(4) \rightarrow \left\{\begin{matrix}
SU(3) \times SU(3) \times U(1) \;,\\
SU(4) \times SU(2) \times SU(2)\times U(1) \;. \\
\end{matrix}\right. 
\end{equation*}

The first choice corresponds to extensively studied models based on $SU(3)_C \times SU(3)_L \times U(1)_N$ \cite{Singer,Pisano,Long,Foot,Montero}.
The second possibility contains the popular Pati-Salam model. 
We shall begin studying the second case, leaving the first one for future works.\\

At this stage we can label one $SU(4)$ as $SU(4)_C$, which will lead to the strong interaction at low energy scales, and the other $SU(4)$ as $SU(4)_{RL}$, which will lead to the left-right symmetry $SU(2)_L \times SU(2)_R$. Notice that we avoid identifying any $SU(4)$ with the electroweak group since the Weinberg-Salam theory of electroweak interactions is not a subgroup of any $SU(4)$ individually, but it will arise from both $SU(4)$ 
. Hence,      
\begin{equation}
G = SU(4)_C \times SU(4)_{RL}\;\left(\times \mathcal{Z}_2\right)\;.
\end{equation}

We consider that we recover the electromagnetic and strong interactions ($G'_{SM}=SU(3)_C \times U(1)_Q $) at low energy scales through the following pattern of symmetry breaking, 
\begin{equation}
G \rightarrow G'_{PS} \rightarrow G_{SM} \rightarrow G'_{SM} \;,
\label{breaking}
\end{equation}
where 
\begin{equation}
G'_{PS} =  SU(4)_C \times SU(2)_L \times SU(2)_R \times U(1)_{\beta}\;,
\label{PSstep}
\end{equation}
and
\begin{equation}
SU(4)_C \rightarrow SU(3)_C \times U(1)_X\;,
\label{su4Cbreak}
\end{equation}
\begin{equation}
SU(4)_{RL} \rightarrow SU(2)_L \times SU(2)_R \times U(1)_\beta\;.
\label{su4RLbreak}
\end{equation}

In (\ref{breaking}) we consider that the $SU(4)_{RL}$ is broken at a higher energy scale than $SU(4)_C$ in order to incorporate a Pati-Salam step ($G'_{PS}$). Nevertheless, this sequence is not a requirement since (\ref{su4Cbreak}) and (\ref{su4RLbreak}) symmetry breakings are independent of each other. Other remark here is that the local symmetry $U(1)_\beta$ could be broken at any scale. The label \textit{RL} refers to groups under which both, left-handed and right-handed fermions, are transformed non-trivially, being the groups lebeled with $R$ and $L$ quiral groups under which left-handed and right-handed are singlets respectively.\\

The unified gauge group $G$ is characterized by two gauge couplings, $g_{4C}$ ($SU(4)_C$) and $g_{4RL}$ ($SU(4)_{RL}$). However, since the factors of the semi simple group are identical, these independent gauge coupling constants can be forced to be equal adding the discrete symmetry $\mathcal{Z}_2$. So at high energy we have a single unified gauge coupling parameter $g_G$ in $G$, 
\begin{equation}
g_G = g_{4C} = g_{4RL} \;.
\end{equation}
In a first breaking stage, the gauge coupling unification is lost and the model has got three gauge coupling parameters in $G'_{PS}$, 
\begin{equation}
g_{4C} (SU(4)_C),\; g_\beta (U(1)_\beta),\; g_{2RL} (SU(2)_L \times SU(2)_R) \;,
\end{equation}
and at low energy scales we recover the three gauge couplings parameters of the Standard Model in $G_{SM}$,
\begin{equation}
g_{3} (SU(3)_C),\; g (SU(2)_L),\; g' (U(1)_Y) \;.
\end{equation}

On another note, the standard model charges arise as linear combination of new hypercharges which are generated in the spontaneous symmetry breaking. Explicitly,
\begin{equation}
SU(2)_R \times U(1)_X \rightarrow U(1)_Y \;,
\label{urxay}
\end{equation}
\begin{equation}
SU(2)_L \times U(1)_Y \rightarrow U(1)_Q\;,
\end{equation}

\section{Group construction}
\label{Groupconstruction}

In this section we shall construct the set of generators for the fundamental representations of both $SU(4)$ groups. Then, we will be able to calculate any representation under a $SU(4)_C \times SU(4)_{RL}$ transformation from this set of generators.\\

To build the generators $T_i$, we need to take into account physical considerations beside the following known mathematical properties, 
\begin{itemize}
\item They are hermitian, $T_j=T_j^\dagger$.
\item They are traceless, $\Trace(T_i)=0$. 
\item They satisfy the usual ortonormalization requirement, $\Trace(T_jT_p)=\frac{1}{2}\delta_{jp}$. 
\end{itemize}

In the first place we study the set of generators of $SU(4)$ of color. We will denote $T^C_i=\dfrac{1}{2}\lambda^C_i$, with $i=1,..,15$; the generators of $SU(4)_C$. The first eight generators can be taken as the generalized generators of $SU(3)_C$, i.e the generalized $4\times4$ Gell-Mann matrices. Therefore, these members of the set of generators are associated with color, the strong SM interaction, and their associated gauge bosons (gluons) will keep massless after all symmetry breakings.
\begin{equation}
\lambda^C_j = \begin{pmatrix}[ccc|c]
& & & 0 \\
& \lambda_{j(3\times3)} & & 0 \\
& & & 0 \\\hline 
 0 & 0 & 0 & 0
\end{pmatrix} \;\;\;, \;\textnormal{for}\;\;j=1,..,8 \;;
\end{equation}
where $\lambda_j$ denote the $3\times3$ Gell-Mann matrices \cite{Gell-Mann}. 
The next six matrices $\{\lambda^C_9..\lambda^C_{14}\}$ can be built in the following common way:
\begin{equation*}
\lambda^C_9 = \begin{pmatrix}[ccc|c]
&  &  & 1 \\ 
& 0_{(3\times3)} &  & 0 \\ 
  &  & & 0 \\ \hline
 1 & 0 & 0 & 0
\end{pmatrix}\;,\;\;\;\;\;\;\;\;\;\lambda^C_{10} = \begin{pmatrix}[ccc|c]
&  &  & -i \\ 
& 0_{(3\times3)} &  & 0 \\ 
  &  & & 0 \\ \hline
 i & 0 & 0 & 0
\end{pmatrix}\;,
\label{genecommon1}
\end{equation*}
\begin{equation*}
\lambda^C_{11} = \begin{pmatrix}[ccc|c]
&  &  & 0 \\ 
& 0_{(3\times3)} &  & 1 \\ 
  &  & & 0 \\ \hline
 0 & 1 & 0 & 0
\end{pmatrix}\;,\;\;\;\;\;\;\;\;\;\lambda^C_{12} = \begin{pmatrix}[ccc|c]
&  &  & 0 \\ 
& 0_{(3\times3)} &  & -i \\ 
  &  & & 0 \\ \hline
 0 & i & 0 & 0
\end{pmatrix}\;,
\end{equation*}
\begin{equation}
\lambda^C_{13} = \begin{pmatrix}[ccc|c]
&  &  & 0 \\ 
& 0_{(3\times3)} &  & 0 \\ 
  &  & &1\\ \hline
 0 & 0 & 1 & 0 
\end{pmatrix} \;,\;\;\;\;\;\;\;\;\;\lambda^C_{14} = \begin{pmatrix}[ccc|c]
&  &  & 0 \\ 
& 0_{(3\times3)} &  & 0 \\ 
  &  & & -i \\ \hline
 0 & 0 & i & 0
\end{pmatrix}\;,
\label{genecommon}
\end{equation}\\
These six generators are associated with \textit{lepto-quarks}, i.e, gauge bosons which rotate quarks into leptons. These processes are forbidden in the Standard Model and they can give rise to interesting phenomenology such as the recent intriguing LHCb flavour anomalies \cite{LHCbanomaly1,LHCbanomaly2,Sahoo}.\\

The last diagonal generator $T^C_{15}$, represents a new hypercharge $X$ which contributes to the electroweak SM hypercharge. Taking into account the ortonormalization requirement we can choose $\lambda^C_{15}$ as  
\begin{equation}
\lambda^C_{15} = \dfrac{\sqrt{6}}{6}\begin{pmatrix} 1 & 0 & 0 & 0 \\
 0 & 1 & 0 & 0 \\
  0 & 0 & 1 & 0 \\
   0 & 0 & 0 & -3
\end{pmatrix}\;. 
\end{equation}\\
The set of generators constructed $\{T_i^C\}$ allows us to observe a clear structure in $SU(4)_C$:
\begin{equation*}
\begin{pmatrix}  &  &   \\
  & SU(4)_C  &    \\
   & & 
\end{pmatrix} = \left(\begin{array}{ccc|c}
\cellcolor{green!50} & \cellcolor{green!50} & \cellcolor{green!50} & \cellcolor{yellow} \textnormal{lepto-} \\ 
\cellcolor{green!50}& SU(3)_C \cellcolor{green!50}& \cellcolor{green!50} & \cellcolor{yellow} \textnormal{quarks} \\ 
 \cellcolor{green!50} & \cellcolor{green!50} & \cellcolor{green!50}& \cellcolor{yellow} \\ \hline
\cellcolor{yellow} & \cellcolor{yellow} \textnormal{lepto-quarks}& \cellcolor{yellow}  & \cellcolor{purple!50} U(1)_X 
\end{array}\right). 
\end{equation*}\\
This structure will be more apparent when we build the gauge bosons matrices in section \ref{Gaugebosons}.\\

Now we consider the other $SU(4)$. We shall denote $T^{RL}_i=\dfrac{1}{2}\lambda^{RL}_i$, with $i=1,..,15$; the generators of $SU(4)_{RL}$. In order to get a pattern of symmetry breaking as (\ref{breaking}) we can take:
\begin{equation*}
\lambda^{RL}_j = \begin{pmatrix}[c|c]
 \sigma_{j(2\times2)}  & 0_{(2\times2)} \\\hline 
  0_{(2\times2)} & 0_{(2\times2)}
\end{pmatrix} \;\;\;\;, \;\textnormal{for}\;\;j=1,2,3\;;
\label{su2L}
\end{equation*}
\begin{equation*}
\lambda^{RL}_j = \begin{pmatrix}[c|c]
 0_{(2\times2)} & 0_{(2\times2)} \\\hline 
 0_{(2\times2)} & \sigma_{j-3(2\times2)}
\end{pmatrix} \;, \;\textnormal{for}\;\;j=4,5,6\;;
\label{su2R}
\end{equation*}\\
where $\sigma_j$ are the Pauli matrices. Each subset of generators ($j=1,2,3$ and $j=4,5,6$) corresponds to the subgroups $SU(2)_{L}$ and $SU(2)_{R}$ respectively.
Hence, the gauge bosons associated with these generators keep massless at high energy scales, after the first symmetry breaking. However, the $SU(2)_R$ bosons have to be more massive than the $SU(2)_L$ bosons since they acquire their mass before the electroweak Higgs gets its vacuum expectation value (VEV).  It will be useful to display explicitly the diagonal generators:
\begin{equation}
\lambda^{RL}_{3} = \begin{pmatrix}
1 & 0 & 0 & 0 \\
 0 & -1 & 0 & 0 \\
  0 & 0 & 0 & 0 \\
   0 & 0 & 0 & 0
\end{pmatrix}\;,
\end{equation}
\begin{equation}
\lambda^{RL}_{6} = \begin{pmatrix}
0 & 0 & 0 & 0 \\
 0 & 0 & 0 & 0 \\
  0 & 0 & 1 & 0 \\
   0 & 0 & 0 & -1
\end{pmatrix}\;.
\end{equation}\\
The next eight generators of $SU(4)_{RL}$, $\lambda^{RL}_i$ for $i=7,..,14$, can be constructed in the same way we did for $SU(4)_C$ (\ref{genecommon}): 
\begin{equation*}
\lambda^{RL}_7 = \begin{pmatrix}[cc|cc]
&  & 1 & 0 \\ 
 \multicolumn{2}{c|}{\smash{\raisebox{.5\normalbaselineskip}{$0_{(2\times2)}$}}} & 0 & 0 \\ \hline 
 1 & 0 & & \\ 
 0 & 0 & \multicolumn{2}{c}{\smash{\raisebox{.5\normalbaselineskip}{$0_{(2\times2)}$}}}
\end{pmatrix}\;,\;\;\;\;\;\;\;\;\;\lambda^{RL}_8 = \begin{pmatrix}[cc|cc]
&  & -i & 0 \\ 
 \multicolumn{2}{c|}{\smash{\raisebox{.5\normalbaselineskip}{$0_{(2\times2)}$}}} & 0 & 0 \\ \hline 
 i & 0 & & \\ 
 0 & 0 & \multicolumn{2}{c}{\smash{\raisebox{.5\normalbaselineskip}{$0_{(2\times2)}$}}}
\end{pmatrix}\;,
\label{genecommonRL1}
\end{equation*}
\begin{equation*}
\lambda^{RL}_9 = \begin{pmatrix}[cc|cc]
&  & 0 & 1 \\ 
 \multicolumn{2}{c|}{\smash{\raisebox{.5\normalbaselineskip}{$0_{(2\times2)}$}}} & 0 & 0 \\ \hline 
 0 & 0 & & \\ 
 1 & 0 & \multicolumn{2}{c}{\smash{\raisebox{.5\normalbaselineskip}{$0_{(2\times2)}$}}}
\end{pmatrix}\;,\;\;\;\;\;\;\;\;\;\lambda^{RL}_{10} = \begin{pmatrix}[cc|cc]
&  & 0 & -i \\ 
 \multicolumn{2}{c|}{\smash{\raisebox{.5\normalbaselineskip}{$0_{(2\times2)}$}}} & 0 & 0 \\ \hline 
 0 & 0 & & \\ 
 i & 0 & \multicolumn{2}{c}{\smash{\raisebox{.5\normalbaselineskip}{$0_{(2\times2)}$}}}
\end{pmatrix}\;,
\end{equation*}
\begin{equation*}
\lambda^{RL}_{11} = \begin{pmatrix}[cc|cc]
&  & 0 & 0 \\ 
 \multicolumn{2}{c|}{\smash{\raisebox{.5\normalbaselineskip}{$0_{(2\times2)}$}}} & 1 & 0 \\ \hline 
 0 & 1 & & \\ 
 0 & 0 & \multicolumn{2}{c}{\smash{\raisebox{.5\normalbaselineskip}{$0_{(2\times2)}$}}}
\end{pmatrix} \;,\;\;\;\;\;\;\;\;\;\lambda^{RL}_{12} = \begin{pmatrix}[cc|cc]
&  & 0 & 0 \\ 
 \multicolumn{2}{c|}{\smash{\raisebox{.5\normalbaselineskip}{$0_{(2\times2)}$}}} & -i & 0 \\ \hline 
 0 & i & & \\ 
 0 & 0 & \multicolumn{2}{c}{\smash{\raisebox{.5\normalbaselineskip}{$0_{(2\times2)}$}}}
\end{pmatrix}\;,
\label{genecommonRL2}
\end{equation*}
\begin{equation}
\lambda^{RL}_{13} = \begin{pmatrix}[cc|cc]
&  & 0 & 0 \\ 
 \multicolumn{2}{c|}{\smash{\raisebox{.5\normalbaselineskip}{$0_{(2\times2)}$}}} & 0 & 1 \\ \hline 
 0 & 0 & & \\ 
 0 & 1 & \multicolumn{2}{c}{\smash{\raisebox{.5\normalbaselineskip}{$0_{(2\times2)}$}}}
\end{pmatrix} \;,\;\;\;\;\;\;\;\;\;\lambda^{RL}_{14} = \begin{pmatrix}[cc|cc]
&  & 0 & 0 \\ 
 \multicolumn{2}{c|}{\smash{\raisebox{.5\normalbaselineskip}{$0_{(2\times2)}$}}} & 0 & -i \\ \hline 
 0 & 0 & & \\ 
 0 & i & \multicolumn{2}{c}{\smash{\raisebox{.5\normalbaselineskip}{$0_{(2\times2)}$}}}
\end{pmatrix}\;,
\label{genecommonRL3}
\end{equation}\\
The gauge bosons associated to these eight generators will be the heaviest bosons since they acquire their mass in the first symmetry breaking.\\ 

The last diagonal generator $\lambda^{RL}_{15}$ is not unique after the choice of $\lambda^{RL}_{3}$ and $\lambda^{RL}_{6}$, although it is constrained due to the ortonormalization requirement. For convenience we define it as follows:
\begin{equation}
\lambda^{RL}_{15} =\dfrac{\sqrt{2}}{2}\begin{pmatrix}
1 & 0 & 0 & 0 \\
 0 & 1 & 0 & 0 \\
  0 & 0 & -1 & 0 \\
   0 & 0 & 0 & -1
\end{pmatrix} \;.
\end{equation}
As in the case of $SU(4)_C$, the group $SU(4)_{RL}$ shows a direct structure using this set of generators $\{T_i^{RL}\}$:     
\begin{equation*}
\begin{pmatrix}
  &  &  \\
   & SU(4)_{RL}   &  \\
    &  & 
\end{pmatrix} = \left(\begin{array}{cc|cc}
\cellcolor{orange!80}&\cellcolor{orange!80} SU(2)_L & \cellcolor{yellow} & \cellcolor{yellow} \textnormal{heavy} \\ 
\cellcolor{orange!80}&\cellcolor{orange!80} & \cellcolor{yellow} & \cellcolor{yellow} \textnormal{bosons} \\\hline 
 \cellcolor{yellow} & \cellcolor{yellow}\textnormal{heavy} & \cellcolor{blue!30} & \cellcolor{blue!30} \\ 
\cellcolor{yellow} & \cellcolor{yellow} \textnormal{bosons} & \cellcolor{blue!30}  & \cellcolor{blue!30} SU(2)_R 
\end{array}\right) \;.
\end{equation*}

All displayed generators satisfy the algebra of $SU(4)_{C}\times SU(4)_{RL}$  given by:
\begin{equation}
\comm{T^{i}_a}{T^{j}_b}=i\delta^{ij}\sum_c f^i_{abc}T^i_c \;\;;
\label{conmut}
\end{equation}
where $i,j=C, RL$ and $f_{abc}$ are the structure constants of $SU(4)$ (see Appendix \ref{structurefuncappen}).\\ 

From the set of generators $\left\{T_i\right\}$ for the fundamental representations $4$ constructed, we can calculate any representation under a $SU(4)$ transformation. In particular, we can calculate the quantum numbers (see section \ref{quanumbersection}), which are determined by the diagonal generators, of the lowest-dimensional $SU(4)$ representations from the defined generators as follows: 
\begin{itemize}
\item The anti-fundamental representation $\bar{4}$,
\begin{equation}
\bar{T}_i^{(ab)} = - T_i^{(ab)}\delta_{ab}
\end{equation}
\item The anti-symmetric representation $6$,
\begin{equation}
T^{6(ab)}_i = T_i^{(aa)} + T_i^{(bb)} - 2 T_i^{(aa)}\delta_{ab}
\end{equation}
\item The symmetric representation $10$,
\begin{equation}
T^{10(ab)}_i = -T_i^{(aa)} - T_i^{(bb)}
\end{equation}
\item The adjoint representation $15$,
\begin{equation}
T^{15(ab)}_i = T_i^{(aa)} - T_i^{(bb)}
\end{equation}
\end{itemize}
where $T_i$ are the diagonal generators of the fundamental representation and $a,b = 1,2,3,4$ are matrix indices. Here and in section \ref{quanumbersection}, we are employing a notation in which the matrix elements are the associated eigenvalues, i.e the diagonal elements of the generator of the quantum number for each representation.  

\section{Gauge bosons}
\label{Gaugebosons}

Our model has thirty gauge bosons since each $SU(4)$ has fifteen ($4^2-1$) generators. We remind that the Standard Model has got twelve gauge bosons and the Pati-Salam model twenty-one. Therefore, Double $SU(4)$ GUT involves new bosons which acquire mass at different energy scales.\\

By using the generators constructed in the previous section, we can build the convenient $4\times4$ matrix of gauge bosons $A$ defined as:
\begin{equation}
A= \sum_{i=1}^{15}T_iA^i \;,
\end{equation}
where $A^i\equiv A_{\mu}^i$ are vector fields. Thus, in the case of $SU(4)_C$:

\begin{equation}
A_C=\dfrac{1}{\sqrt{2}}\begin{pmatrix}
\frac{\sqrt{12}}{12}Y'+\frac{\sqrt{6}}{6}G^8+\frac{\sqrt{2}}{2}G^3 & G^1 & G^4 & \bar{X}^1 \\
G^2 & \frac{\sqrt{12}}{12}Y'+\frac{\sqrt{6}}{6}G^8 - \frac{\sqrt{2}}{2}G^3  &  G^6 & \bar{X}^2 \\
G^5 & G^7 & \frac{\sqrt{12}}{12}Y'-\frac{\sqrt{6}}{3}G^8 &  \bar{X}^3\\
 X^1 & X^2  & X^3 &  -\frac{\sqrt{12}}{4}Y'
\end{pmatrix}\;,
\end{equation} 
where $G^\alpha\equiv G^\alpha_\mu$ with $\alpha=1,2,...,8$; are the QCD gluons,
\begin{equation}
G^{1|2}_\mu=\frac{1}{\sqrt{2}}\left(A^1_\mu \pm i A^2_\mu\right) \;,
\end{equation}
\begin{equation}
G^{4|5}_\mu=\frac{1}{\sqrt{2}}\left(A^4_\mu \pm i A^5_\mu\right) \;,
\end{equation}
\begin{equation}
G^{6|7}_\mu=\frac{1}{\sqrt{2}}\left(A^6_\mu \pm i A^7_\mu\right) \;,
\end{equation}
\begin{equation}
G^{3}_\mu= A^3_\mu\;,\;\;\;G^{8}_\mu= A^8_\mu \;,
\end{equation}
$X^\gamma\equiv X^\gamma_\mu$ and $\bar{X}^\gamma\equiv \bar{X}^\gamma_\mu$ with $\gamma=$1, 2 and 3; are the lepto-quarks,
\begin{equation}
X^{1}_\mu,\bar{X}^{1}_\mu=\frac{1}{\sqrt{2}}\left(A^9_\mu \pm i A^{10}_\mu\right) \;,
\end{equation}
\begin{equation}
X^{2}_\mu,\bar{X}^{2}_\mu=\frac{1}{\sqrt{2}}\left(A^{11}_\mu \pm i A^{12}_\mu\right) \;,
\end{equation}
\begin{equation}
X^{3}_\mu,\bar{X}^{3}_\mu=\frac{1}{\sqrt{2}}\left(A^{13}_\mu \pm i A^{14}_\mu\right) \;,
\end{equation}
and $Y'\equiv Y_\mu' = A_\mu^{15}$ is the hyperphoton associated with the intermediate local Abelian symmetry $U(1)_{X}$.\\

Those fifteen gauge bosons $\left(G^\alpha\right.$, $X^\gamma$, $\bar{X}^\gamma$, $\left.Y'\right)$ are transformed according to the representation $(15,1)$ of $SU(4)_C \times SU(4)_{RL}$, i.e, according to the adjoint representation of $SU(4)_C$, and invariant under a $SU(4)_{RL}$ transformation. The adjoint representation decomposes into the following irreducible representations of $SU(4)_C \rightarrow SU(3)_C \times U(1)_{X}$:
\begin{equation}
15 \;\rightarrow \;8_0 \;\;\oplus\;\; 3_{\frac{4}{3}} \;\;\oplus\;\; \bar{3}_{-\frac{4}{3}} \;\;\oplus\;\; 1_0   
\end{equation}
Hereafter, we use the lower indices for $U(1)$ charges and the bigger labels for the dimension of representations of non-Abelian groups. The eight gluons are transformed as the adjoint representation of the group $SU(3)_C$, therefore they correspond to $8_0$. The hyperphoton $Y'$ is transformed as a singlet of color $1_0$. Thus we observe that the lepto-quarks are triplets and anti-triplets under a color transformation, $3_{\frac{4}{3}}$ and $\bar{3}_{-\frac{4}{3}}$ respectively.\\

For $SU(4)_{RL}$ we obtain the following matrix of gauge bosons:

\begin{equation}
A_{RL}=\dfrac{1}{\sqrt{2}}\begin{pmatrix}
\frac{\sqrt{2}}{2}W_L^3 + \frac{1}{2}\beta' & W_L^+ & \bar{Y^1} & \bar{Y^2} \\
W_L^- & -\frac{\sqrt{2}}{2}W_L^3 + \frac{1}{2}\beta' & \bar{Y^3} & \bar{Y^4}  \\
 Y^1 & Y^3 & \frac{\sqrt{2}}{2}W_R^3 - \frac{1}{2}\beta' & W_R^+ \\
 Y^2 & Y^4 & W_R^- &  -\frac{\sqrt{2}}{2}W_R^3 - \frac{1}{2}\beta'
\end{pmatrix}\;,
\end{equation}
where $W_L^{\alpha}$ $\left(\alpha=+,-,3\right)$ are the vector bosons of the gauge group $SU(2)_L$ of the electroweak interactions present in the Standard Model ($G_{SM}$),
\begin{equation}
W_{L}^{+|-}=\frac{1}{\sqrt{2}}\left(A^1_\mu \pm i A^2_\mu\right) \;, \;\;W_L^3=A_\mu^3\;,
\end{equation}
The bosons $W_R^{\kappa}$ $\left(\kappa=+,-,3\right)$ are those of the gauge group $SU(2)_R$ associated with the left-right symmetry at intermediate energy scales, being  
\begin{equation}
W_R^{+|-}=\frac{1}{\sqrt{2}}\left(A^4_\mu \pm i A^5_\mu\right) \;,\;\;W_R^3=A_\mu^6\;,
\end{equation}
$Y^\gamma\equiv Y^\gamma_\mu$ and $\bar{Y}^\gamma\equiv \bar{Y}^\gamma_\mu$ with $\gamma=$1, 2, 3 and 4; are the heaviest bosons which acquire mass in the first symmetry breaking stage, 
\begin{equation}
Y^{1},\bar{Y}^{1}=\frac{1}{\sqrt{2}}\left(A^7_\mu \pm i A^{8}_\mu\right) \;,
\end{equation}
\begin{equation}
Y^{2},\bar{Y}^{2}=\frac{1}{\sqrt{2}}\left(A^{9}_\mu \pm i A^{10}_\mu\right) \;,
\end{equation}
\begin{equation}
Y^{3},\bar{Y}^{3}=\frac{1}{\sqrt{2}}\left(A^{11}_\mu \pm i A^{12}_\mu\right) \;,
\end{equation}
\begin{equation}
Y^{4},\bar{Y}^{4}=\frac{1}{\sqrt{2}}\left(A^{13}_\mu \pm i A^{14}_\mu\right) \;,
\end{equation}\\
and $\beta'\equiv \beta'_\mu = A_\mu^{15}$ is a new hyperphoton which mediates the $U(1)_{\beta}$ local symmetry.\\

The fifteen gauge bosons fields $\left(W_L^\alpha\right.$, $W_R^\kappa$, $Y^\gamma$, $\bar{Y}^\gamma$, $\left.\beta'\right)$ are transformed according to the representation $(1,15)$ of $SU(4)_C \times SU(4)_{RL}$ i.e, according to the adjoint representation of $SU(4)_{RL}$, and invariant under a $SU(4)_{C}$ transformation. The adjoint representation decomposes into the following irreducible representations of $SU(4)_{RL}\rightarrow SU(2)_L \times SU(2)_{R} \times U(1)_\beta$:
\begin{equation}
\begin{split}
15 \rightarrow & \;\; (3,1)_{0} \;\oplus \;(1,3)_{0}\; \oplus\;(1,1)_{0}\; \oplus \; (2,2)_{2} \; \oplus \; (2,2)_{-2} 
\end{split}
\label{repRL1}
\end{equation}
and into the following representations of $SU(2)_L \times SU(2)_{R} \rightarrow SU(2)_L \times U(1)_R$:
\begin{equation}
\begin{split}
 (3 &,1)\;\oplus \;(1,3)\; \oplus\;(1,1)\;\; \oplus \; (2,2) \oplus \; (2,2) \;\; \rightarrow \;\; \\ \rightarrow & \;\; \left(3_0\right) \; \oplus \;  \left(1_1 \; \oplus \; 1_0 \; \oplus \; 1_{-1}\right) \; \oplus \; \left(1_0\right) \\ & \; \oplus \; \left(2_{\frac{1}{2}} \; \oplus \; 2_{-\frac{1}{2}}\right) \; \oplus \; \left(2_{\frac{1}{2}} \; \oplus \; 2_{-\frac{1}{2}}\right)  
\end{split}
\label{repRL2}
\end{equation}
Therefore, we can identify the representations of the Standard Model gauge group ($G_{SM}$) in which each gauge boson is at low energies. We display them on Table \textcolor{purple}{1}. We have used the extended Gell-Mann Okubo relation of the model derived in section \ref{quanumbersection} to calculate their Standard Model weak hypercharge values.\\

The gauge invariant kinetic energy terms of the Double SU(4) GUT Lagrangian for the gauge bosons are:
\begin{equation}
\mathcal{L}_G = -\frac{1}{2}\Tr\left(C_{\mu\nu}C^{\mu\nu}\right) - \frac{1}{2}\Tr\left(W_{\mu\nu}W^{\mu\nu}\right)  , 
\end{equation}
where,
\begin{equation}
C_{\mu\nu} \equiv \partial_\mu \left(A_C\right)_\nu - \partial_\nu \left(A_C\right)_\mu -ig_{4C} \left[\left(A_C\right)_\mu , \left(A_C\right)_\nu\right] \;,
\end{equation}
\begin{equation}
W_{\mu\nu} \equiv \partial_\mu \left(A_{RL}\right)_\nu - \partial_\nu \left(A_{RL}\right)_\mu - i g_{4RL} \left[\left(A_{RL}\right)_\mu , \left(A_{RL}\right)_\nu\right] \;.
\end{equation}

\begin{table*}
\scriptsize\centering
\begin{tabular}{|c||c|c|c|} \hline
 Gauge Bosons & $\left(SU(4)_C \times SU(4)_{RL}\right)$ & $\left(SU(4)_C \times SU(2)_L \times SU(2)_R\right)$ & $\left(SU(3)_C \times SU(2)_L \right)_{U(1)_Y}$  \\\hline
 \multirow{2}{*}{$SU(4)_C$}  & \multirow{2}{*}{(15,1)} & \multirow{2}{*}{(15,1,1)} & $G^{\alpha} \sim (8,1)_0  \oplus  X^\gamma \sim (3,1)_{-\frac{4}{3}}$ \\
  & & &  $\bar{X}^\gamma \sim (\bar{3},1)_{\frac{4}{3}}\oplus (1,1)_0$ \\\hline
 \multirow{6}{*}{$SU(4)_{RL}$} & \multirow{6}{*}{(1,15)} & (1,3,1) & $W_L^\alpha\sim(1,3)_0$ \\\cline{3-4}
 & & \multirow{2}{*}{(1,1,3)} & $W_R^+\sim(1,1)_2 \oplus (1,1)_{0}$ \\
 & & & $\oplus \;W_R^-\sim(1,1)_{-2}$ \\\cline{3-4}
  & & (1,1,1) & $\beta' \sim (1,1)_0$ \\\cline{3-4}
  & & \multirow{2}{*}{2$\cdot$(1,2,2)} & $Y^{1,3}\sim(1,2)_{1}\oplus Y^{2,4}\sim(1,2)_{-1}$\\
  & & & $\bar{Y}^{1,3}\sim(1,2)_{-1}\oplus \bar{Y}^{2,4}\sim(1,2)_{1}$ \\\hline
\end{tabular}
\label{br}
\caption{Gauge bosons representations at different energy scales. Notice that there are two singlets $(1,1)_0$ without associated bosons. They correspond to a massless lineal combination of $Y'$ and $W_R^3$, the weak hyperfoton $B$ (\ref{urxay}), and to other massive boson combination of $Y'$ and $W_R^3$, both generated in the spontaneous symmetry breaking.} 
\end{table*}

\section{Quantum numbers}
\label{quanumbersection}

In this section, we introduce the quantum numbers of the Double SU(4) GUT. Furthermore, we derive the extended Gell-Mann Okubo relation in the framework under consideration from these quantum numbers.\\

The rank of the unified gauge group $G$ is six, since there are six diagonal generators or Cartan generators (three for each $SU(4)$). Thereby, at high energy, there are six different conserved quantum numbers that characterize the particles in the model. These quantum numbers have to generate the Pati-Salam charges at an intermediate energy scale and the Standard Model charges at low energies. One of the advantages of Grand Unification is to give a natural explanation for the U(1) charge quantization through the spontaneous symmetry breaking. It is convenient to define the Cartan subalgebra of $SU(4)_{RL}$ and $SU(4)_C$ composed of \{$T^{RL}_{3}$, $T^{RL}_{6}$, $T^{RL}_{15}$\} and \{$T^{C}_{3}$, $T^{C}_{8}$, $T^{C}_{15}$\} respectively. Thus, the linear combinations of these six generators will give rise to all quantum numbers in the model. \\ 

The quantum numbers built by \{$T^{C}_{3}$, $T^{C}_{8}$\} will be associated to the subgroup $SU(3)_C$, the color group of the Standard Model, and hence they are the two chromodynamics quantum numbers. Since the $SU(3)_C$ symmetry is conserved even at low energies, there are not new physical implications related to them. The additional quantum number of $SU(4)_C$ corresponds to the hypercharge of the unbroken $U(1)_{X}$, so we can define it in function of the remaining Cartan generator of $SU(4)_C$ as it is usual in the Pati-Salam model.
\begin{itemize}
\item X hypercharge:
\begin{equation}
X \equiv B-L = \dfrac{2\sqrt{6}}{3}T^{C}_{15}\;,
\label{X}
\end{equation}
\end{itemize}
where $B$ is the baryon number, and $L$ is the lepton number.\\ 

The other quantum numbers are constructed by the Cartan generators of $SU(4)_{RL}$, therefore they are associated with the electroweak sector of the Standard Model (besides $X$). We have several possibilities to combine linearly the diagonal generators in order to determined independent quantum numbers. We define them in the following convenient way.
\begin{itemize}
\item Weak 4-isospin:
\begin{equation}
I_4 = T^{RL}_3 + T^{RL}_6 + 2\sqrt{2}T^{RL}_{15}\;.
\end{equation}
\item Right-left weak Isospin:
\begin{equation}
I_3^{RL} = T^{RL}_3 - T^{RL}_6\;.
\end{equation}
\item Beta hypercharge:
\begin{equation}
\beta = 2\sqrt{2}T^{RL}_{15}\;.
\end{equation}
\end{itemize}
The \textit{weak 4-isospin} allows us to label the components of a $SU(4)_{RL}$ multiplet which is transformed as the fundamental representation, therefore a four-dimensional representation, with $I_4= 3/2, 1/2, -1/2, -3/2$ as it is common for a multiplet of spin $3/2$ with four components. The \textit{Right-left weak Isospin} labels the components of the two-dimensional broken multiplets at the Pati-Salam stage with $I_3^{RL}=1/2, -1/2$. Lastly, the \textit{Beta hypercharge} is a quantum number which assigns +1 and -1 to different degrees of freedom in the fundamental and anti-fundamental representations of $SU(4)_{RL}$. It is associated with the unbroken continuous symmetry $U(1)_\beta$.\\

We can build all charges of the different symmetry breaking stages by linear combinations of these quantum numbers. The Pati-Salam model charges, which are conserved until intermediate energy scales in the left-right group symmetry $SU(2)_L \times SU(2)_R$ (\ref{PSstep}), are 
\begin{equation}
I_3^{R} =\dfrac{1}{2} \left(I_4 - I_3^{RL} - \beta \right) \;, 
\end{equation}
\begin{equation}
I_3^{L} =\dfrac{1}{2} \left(I_4 + I_3^{RL} - \beta \right) \;.
\end{equation}\\
The SM hypercharge is recovered at the electroweak scale, being the following combination of the defined quantum numbers:
\begin{equation}
Y = I_4 - I_3^{RL} - \beta + X \;.
\end{equation}

From these results, we can finally derive the extended Gell-Mann Okubo relation in the framework of Double SU(4) GUT considered. The electromagnetic charge, present at low energies after the electroweak Higgs acquire its VEV, is given by: 
\begin{equation}
Q= I_3^L + \dfrac{Y}{2} = I_4 - \beta + X/2 \;.
\end{equation}
This is the extended Gell-Mann Okubo relation for our model.
Explicitly, the electric charge operator in the four-dimensional fundamental representation of $SU(4)_C$ (singlet under a $SU(4)_{RL}$ transformation) is 
\begin{equation}
Q(4_{C},1_{RL}) = \begin{pmatrix}
\frac{1}{6} & 0 & 0 & 0 \\
0 & \frac{1}{6} & 0 & 0 \\
0 & 0 & \frac{1}{6} & 0 \\
0 & 0 & 0 & -\frac{1}{2}
\end{pmatrix}\;,
\end{equation} 
and in the fundamental representation of $SU(4)_{RL}$ (singlet under $SU(4)_{C}$ transformations) is
\begin{equation}
Q(1_C,4_{RL}) = \begin{pmatrix}
\frac{1}{2} & 0 & 0 & 0 \\
0 & -\frac{1}{2} & 0 & 0 \\
0 & 0 & \frac{1}{2} & 0 \\
0 & 0 & 0 & -\frac{1}{2}
\end{pmatrix}\;.
\end{equation} 
Thus, the charge operator in the fundamental $(4,4)$ representation of $SU(4)_c\times SU(4)_{RL}$ can be written as
\begin{equation}
Q(4_C,4_{RL})= \begin{pmatrix}
\frac{2}{3} & \frac{2}{3} & \frac{2}{3} & 0 \\
-\frac{1}{3} & -\frac{1}{3} & -\frac{1}{3} & -1\\
\frac{2}{3} & \frac{2}{3} & \frac{2}{3} & 0\\
-\frac{1}{3} & -\frac{1}{3} & -\frac{1}{3} & -1
\end{pmatrix} \;,
\end{equation}
where we are employing a notation in which $SU(4)_C$ indices run horizontally and $SU(4)_{RL}$ indices run vertically.\\

In Tables \textcolor{purple}{2}, \textcolor{purple}{3}, \textcolor{purple}{4} and \textcolor{purple}{5}, we display all quantum numbers and SM charges of the Double $SU(4)$ GUT particles.

\begin{table*}[h!]
\small\centering
\begin{tabular}{|c|c|c|c|c|c|c|c|c|c|c|c|c|c|c|c|c|} \hline
 & $G^1$ & $G^2$ & $G^3$ & $G^4$ & $G^5$ & $G^6$ & $G^7$ & $G^8$ & $X^1$ & $X^2$ & $X^3$ & $\bar{X}^1$ & $\bar{X}^2$ & $\bar{X}^3$ & $Y'$   \\\hline
 X & 0 & 0 & 0 & 0 & 0 & 0 & 0 & 0 & $-\frac{4}{3}$ & $-\frac{4}{3}$ & $-\frac{4}{3}$ & $\frac{4}{3}$ & $\frac{4}{3}$ & $\frac{4}{3}$ & 0 \\\hline
 I$_3^L$ & 0 & 0 & 0 & 0 & 0 & 0 & 0 & 0 & 0 & 0 & 0 & 0 & 0 & 0 & 0 \\\hline
 Y & 0 & 0 & 0 & 0 & 0 & 0 & 0 & 0 & $-\frac{4}{3}$ & $-\frac{4}{3}$ & $-\frac{4}{3}$ & $\frac{4}{3}$ & $\frac{4}{3}$ & $\frac{4}{3}$ & 0 \\\hline
 Q & 0 & 0 & 0 & 0 & 0 & 0 & 0 & 0 & $-\frac{2}{3}$ & $-\frac{2}{3}$ & $-\frac{2}{3}$ & $\frac{2}{3}$ & $\frac{2}{3}$ & $\frac{2}{3}$ & 0 \\\hline
\end{tabular}
\label{bosonsquantumnumbersC}
\caption{Quantum numbers and SM charges of $SU(4)_{C}$ gauge bosons present in the Double SU(4) GUT.}
\end{table*}
\begin{table*}[h!]
\small\centering
\begin{tabular}{|c|c|c|c|c|c|c|c|c|c|c|c|c|c|c|c|c|} \hline
 & $W_L^+$ & $W_L^-$ & $W_L^3$ & $W_R^+$ & $W_R^-$ & $W_L^3$ & $Y^1$ & $Y^2$ & $Y^3$ & $Y^4$ & $\bar{Y}^1$ & $\bar{Y}^2$ & $\bar{Y}^3$ & $\bar{Y}^4$ & $\beta'$   \\\hline
 I$_4$ & 1 & -1 & 0 & 1 & -1 & 0 & -2 & -3 & -1 & -2 & 2 & 3 & 1 & 2 & 0 \\\hline
 $\beta$  & 0 & 0 & 0 & 0 & 0 & 0 & -2 & -2 & -2 & -2 & 2 & 2 & 2 & 2 & 0 \\\hline
 I$^{RL}_3$ & 1 & -1 & 0 & -1 & 1 & 0 & -1 & 0 & 0 & 1 & 1 & 0 & 0 & -1 & 0\\\hline
 I$_3^L$ & 1 & -1 & 0 & 0 & 0 & 0 & $-\frac{1}{2}$ & $-\frac{1}{2}$ & $\frac{1}{2}$ & $\frac{1}{2}$ & $\frac{1}{2}$ & $\frac{1}{2}$ & -$\frac{1}{2}$ & -$\frac{1}{2}$ & 0\\\hline
 Y & 0 & 0 & 0 & 2 & -2 & 0 & 1 & -1 & 1 & -1 & -1 & 1 & -1 & 1 & 0 \\\hline
 Q & 1 & -1 & 0 & 1 & -1 & 0 & 0 & -1 & 1 & 0 & 0 & 1 & -1 & 0 & 0  \\\hline
\end{tabular}
\label{bosonsquantumnumbersRL}
\caption{Quantum numbers and SM charges of $SU(4)_{RL}$ gauge bosons present in the Double SU(4) GUT.}
\end{table*}
\begin{table*}[h!]
\scriptsize\centering
\begin{tabular}{|c|c|c|c|c|c|c|c|c|c|c|c|c|c|c|c|c|}\hline
 & u & d & e$^-$ & $\nu$ & U & D & E$^-$ & V & u$^c$ & d$^c$ & e$^+$ & $\nu^c$ & U$^c$ & D$^c$ & E$^+$ & V$^c$   \\\hline
 I$_4$ & $\frac{3}{2}$ & $\frac{1}{2}$ & $\frac{1}{2}$ & $\frac{3}{2}$ & $-\frac{1}{2}$ & $-\frac{3}{2}$ & $-\frac{3}{2}$ & $-\frac{1}{2}$ & $\frac{1}{2}$ & $\frac{3}{2}$ & $\frac{3}{2}$ & $\frac{1}{2}$ & $-\frac{3}{2}$ & $-\frac{1}{2}$ & $-\frac{1}{2}$  & $-\frac{3}{2}$ \\\hline
 X& $\frac{1}{3}$ & $\frac{1}{3}$ & $-1$ & $-1$ & $\frac{1}{3}$ & $\frac{1}{3}$ & $-1$ & $-1$  & $-\frac{1}{3}$ & $-\frac{1}{3}$ & 1 & 1 & $-\frac{1}{3}$ & $-\frac{1}{3}$ & 1 & 1  \\\hline
 I$^{RL}_3$ & $\frac{1}{2}$ & $-\frac{1}{2}$ & $-\frac{1}{2}$ & $\frac{1}{2}$ & $-\frac{1}{2}$ & $\frac{1}{2}$ & $\frac{1}{2}$ & $-\frac{1}{2}$ & $\frac{1}{2}$ & $-\frac{1}{2}$ & $-\frac{1}{2}$ & $\frac{1}{2}$ & $-\frac{1}{2}$ & $\frac{1}{2}$ & $\frac{1}{2}$ & $-\frac{1}{2}$ \\\hline
 $\beta$  & 1 & 1 & 1 & 1 & -1 & -1 & -1 & -1 & 1 & 1 & 1 & 1 & -1 & -1 & -1 & -1 \\\hline
 I$^L_3$ &  $\frac{1}{2}$  &  $-\frac{1}{2}$  &  $-\frac{1}{2}$ & $\frac{1}{2}$ & 0 & 0 & 0 & 0 & 0 & 0 & 0 & 0 &  $-\frac{1}{2}$ &  $\frac{1}{2}$&  $\frac{1}{2}$ &  $-\frac{1}{2}$ \\\hline
 Y & $\frac{1}{3}$ & $\frac{1}{3}$ & -1 & -1 &  $\frac{4}{3}$ & $-\frac{2}{3}$ & -2 & 0 & $-\frac{4}{3}$ & $\frac{2}{3}$ & 2 & 0 & $-\frac{1}{3}$  & $-\frac{1}{3}$ & 1 & 1 \\\hline
 Q & $\frac{2}{3}$ & $-\frac{1}{3}$ & -1 & 0 & $\frac{2}{3}$ & $-\frac{1}{3}$ & -1 & 0 & $-\frac{2}{3}$ & $\frac{1}{3}$ & 1 & 0 &  $-\frac{2}{3}$ & $\frac{1}{3}$ & 1 & 0 \\\hline
\end{tabular}
\label{quantumnumbers}
\caption{Quantum numbers and SM charges of left-handed fermions and antifermions of the first generation present in the $SU(4)_C\times SU(4)_{RL}$ model considered.}
\end{table*}
\begin{table*}[h!]
\scriptsize\centering
\begin{tabular}{|c|c|c|c|c|c|c|c|c|c|c|c|c|c|c|c|c|}\hline
 & u & d & e$^-$ & $\nu$ & U & D & E$^-$ & V & u$^c$ & d$^c$ & e$^+$ & $\nu^c$ & U$^c$ & D$^c$ & E$^+$ & V$^c$   \\\hline
 I$_4$ & $-\frac{1}{2}$ & $-\frac{3}{2}$ & $-\frac{3}{2}$ & $-\frac{1}{2}$ & $\frac{3}{2}$ & $\frac{1}{2}$ & $\frac{1}{2}$ & $\frac{3}{2}$ & $-\frac{3}{2}$ & $-\frac{1}{2}$ & $-\frac{1}{2}$ & $-\frac{3}{2}$ & $\frac{1}{2}$ & $\frac{3}{2}$ & $\frac{3}{2}$  & $\frac{1}{2}$ \\\hline
 X& $\frac{1}{3}$ & $\frac{1}{3}$ & $-1$ & $-1$ & $\frac{1}{3}$ & $\frac{1}{3}$ & $-1$ & $-1$  & $-\frac{1}{3}$ & $-\frac{1}{3}$ & 1 & 1 & $-\frac{1}{3}$ & $-\frac{1}{3}$ & 1 & 1  \\\hline
 I$^{RL}_3$ & $-\frac{1}{2}$ & $\frac{1}{2}$ & $\frac{1}{2}$ & $-\frac{1}{2}$ & $\frac{1}{2}$ & $-\frac{1}{2}$ & $-\frac{1}{2}$ & $\frac{1}{2}$ & $-\frac{1}{2}$ & $\frac{1}{2}$ & $\frac{1}{2}$ & $-\frac{1}{2}$ & $\frac{1}{2}$ & $-\frac{1}{2}$ & $-\frac{1}{2}$ & $\frac{1}{2}$ \\\hline
 $\beta$  & -1 & -1 & -1 & -1 & 1 & 1 & 1 & 1 & -1 & -1 & -1 & -1 & 1 & 1 & 1 & 1 \\\hline
 I$^L_3$ &  0 & 0 & 0 & 0 & $\frac{1}{2}$  &  $-\frac{1}{2}$  &  $-\frac{1}{2}$ & $\frac{1}{2}$  &  $-\frac{1}{2}$ &  $\frac{1}{2}$&  $\frac{1}{2}$ &  $-\frac{1}{2}$ &  0 & 0 & 0 & 0 \\\hline
 Y & $\frac{4}{3}$ & $-\frac{2}{3}$ & -2 & 0 &  $\frac{1}{3}$ & $\frac{1}{3}$ & -1 & -1 & $-\frac{1}{3}$ & $-\frac{1}{3}$ & 1 & 1 & $-\frac{4}{3}$  & $\frac{2}{3}$ & 2 & 0 \\\hline
 Q & $\frac{2}{3}$ & $-\frac{1}{3}$ & -1 & 0 & $\frac{2}{3}$ & $-\frac{1}{3}$ & -1 & 0 & $-\frac{2}{3}$ & $\frac{1}{3}$ & 1 & 0 &  $-\frac{2}{3}$ & $\frac{1}{3}$ & 1 & 0 \\\hline
\end{tabular}
\label{quantumnumbers}
\caption{Quantum numbers and SM charges of right-handed fermions and antifermions of the first generation present in the $SU(4)_C\times SU(4)_{RL}$ model considered.}
\end{table*}

\section{Fermion content}
\label{Fermioncontent}

At this stage, we are ready to assign the particle content appropriately. The 
arrangement of the fermions in the correct multiplets of the gauge group $SU(4)\times SU(4)$, i.e its representation or transformation properties under this group, will determine if our theory is anomaly-free or not \cite{anomalies,anomalies2}. In addition, we should take into account the quantum numbers defined above to obtain the correct charges of the observed particles. We also have to fit the fields in order to get back the Standard Model after the symmetry breaking. Therefore, free anomaly particles assignment can be done fitting the fermions to the multiplets as follows.\\

In the first place, the multiplets of $SU(4)_{RL}$ have to contain the SM $SU(2)_L$-doublets and the $SU(2)_R$-doublets (analogous to $SU(2)_L$-doublets but involving right-handed particles). A straightforward minimal particle assignment would be to fit $SU(2)_L$ and $SU(2)_R$ doublets into a single 4-dimensional multiplet of $SU(4)_{RL}$. Nevertheless, this particle assignment is prohibited due to gauge anomalies associated with triangle graphs of three gauge bosons (see \cite{nuestro}). The appearance of these anomalies rules out the renormazibility of the theory \cite{anomalies}. In order to avoid the anomalies, we may take some of the transformation parameters of the $SU(4)_{RL}$ group as space-time independent. We shall investigate this option in \cite{nuestro}.\\

 Another possibility to cancel the anomalies is to increase the particle spectrum as we shall do in this manuscript. We can construct a vector-like model including exotic degrees of freedom together with currently observed particles in order to complete the multiplets of $SU(4)_{RL}$. Therefore, for one generation, we have   

\begin{equation*}
\begin{pmatrix}
u_\alpha \\
d_\alpha \\
U_\alpha \\
D_\alpha
\end{pmatrix}_L , \;\; 
\begin{pmatrix}
\nu \\
e^- \\
V \\
E^-
\end{pmatrix}_L , \;\; 
\begin{pmatrix}
U_\alpha \\
D_\alpha \\
u_\alpha \\
d_\alpha 
\end{pmatrix}_R , \;\; 
\begin{pmatrix}
V \\
E^- \\
\nu \\
e^- 
\end{pmatrix}_R
\sim 4 \;\;;
\end{equation*}
\begin{equation*}
\begin{pmatrix}
U^c_\alpha \\
D^c_\alpha \\
u^c_\alpha \\
d^c_\alpha 
\end{pmatrix}_L , \;\; 
\begin{pmatrix}
V^c \\
E^+ \\
\nu^c \\
e^+ 
\end{pmatrix}_L , \;\; 
\begin{pmatrix}
u^c_\alpha \\
d^c_\alpha \\
U^c_\alpha \\
D^c_\alpha
\end{pmatrix}_R , \;\; 
\begin{pmatrix}
\nu^c \\
e^+ \\
V^c \\
E^+
\end{pmatrix}_R
\sim \bar{4} \;\;;
\end{equation*}\\
where $\alpha = r, g, b$ are the color indices. Here $u, d$ are the quarks up and down respectively, $u^c$ and $d^c$ are the anti-quarks up and down, $e^-$ and $\nu_e$ are the leptons electron and electronic neutrino, $e^+$ and $\nu_e$ are the anti-leptons positron and electronic anti-neutrino. $U$ and $D$ denote exotic quarks which are transformed as triplets under the strong $SU(3)_C$ interaction, $V$ and $E$ are exotic leptons which are invariant under $SU(3)_C$ transformations, and $U^c$, $D^c$, $V^c$ and $E^+$ are their corresponding charge conjugate state.
$L$ and $R$ represent the projection operators $L\equiv \dfrac{1}{2}(1-\gamma_5)$ and $R\equiv \dfrac{1}{2}(1+\gamma_5)$, i.e, 
$L$ denotes left-handed fermions and $R$ right-handed fermions.\\

On the other side, the multiplets of $SU(4)_C$ are the classic multiplets of the Pati-Salam model, where the leptons correspond to the fourth color. Therefore, the fermions are transformed as the 4-dimensional fundamental representation, and the anti-fermions as the 4-dimensional anti-fundamental representation. For the first generation, we have
\begin{equation*}
\begin{pmatrix}
u_r \\
u_g \\
u_b \\
\nu_e
\end{pmatrix}_L , \;\; 
\begin{pmatrix}
d_r \\
d_g \\
d_b \\
e^-
\end{pmatrix}_L , \;\; 
\begin{pmatrix}
u_r \\
u_g \\
u_b \\
\nu_e
\end{pmatrix}_R , \;\; 
\begin{pmatrix}
d_r \\
d_g \\
d_b \\
e^-
\end{pmatrix}_R 
\sim 4 \;\;;
\end{equation*}
\begin{equation*}
\begin{pmatrix}
U_r \\
U_g \\
U_b \\
V
\end{pmatrix}_L , \;\; 
\begin{pmatrix}
D_r \\
D_g \\
D_b \\
E^-
\end{pmatrix}_L , \;\; 
\begin{pmatrix}
U_r \\
U_g \\
U_b \\
V
\end{pmatrix}_R , \;\; 
\begin{pmatrix}
D_r \\
D_g \\
D_b \\
E^-
\end{pmatrix}_R 
\sim 4 \;\;;
\end{equation*}
\begin{equation*}
\begin{pmatrix}
u^c_r \\
u^c_g \\
u^c_b \\
\nu^c_e
\end{pmatrix}_L , \;\; 
\begin{pmatrix}
d^c_r \\
d^c_g \\
d^c_b \\
e^+
\end{pmatrix}_L , \;\; 
\begin{pmatrix}
u^c_r \\
u^c_g \\
u^c_b \\
\nu^c_e
\end{pmatrix}_R , \;\; 
\begin{pmatrix}
d^c_r \\
d^c_g \\
d^c_b \\
e^+
\end{pmatrix}_R 
\sim \bar{4} \;\;;
\end{equation*}
\begin{equation*}
\begin{pmatrix}
U^c_r \\
U^c_g \\
U^c_b \\
V^c
\end{pmatrix}_L , \;\; 
\begin{pmatrix}
D^c_r \\
D^c_g \\
D^c_b \\
E^+
\end{pmatrix}_L , \;\; 
\begin{pmatrix}
U^c_r \\
U^c_g \\
U^c_b \\
V^c
\end{pmatrix}_R , \;\; 
\begin{pmatrix}
D^c_r \\
D^c_g \\
D^c_b \\
E^+
\end{pmatrix}_R 
\sim \bar{4} \;\;;
\end{equation*}\\

All fields can be written explicitly in a summarized way for each generation as follows. 
 
\begin{itemize}
\item All matter is transformed as $(4,4)$,
\end{itemize}
\begin{equation}
\Psi_{iL} = \begin{pmatrix}
u^i_{r} & u^i_{g} & u^i_{b} & \nu^i \\ 
d^i_{r} & d^i_{g}  & d^i_{b} & e^{-i} \\ 
U^i_{r} & U^i_{g} & U^i_{b} & V^i \\ 
D^i_{r} & D^i_{g}  & D^i_{b} & E^{-i}  
\end{pmatrix}_L\;,\;\;\;
\Psi_{iR} = \begin{pmatrix}
U^i_{r} & U^i_{g} & U^i_{b} & V^i \\ 
D^i_{r} & D^i_{g}  & D^i_{b} & E^{-i}\\
u^i_{r} & u^i_{g} & u^i_{b} & \nu^i \\ 
d^i_{r} & d^i_{g}  & d^i_{b} & e^{-i} \\   
\end{pmatrix}_R \;,
\label{fermion}
\end{equation}
\begin{itemize}
\item All anti-matter is transformed as $(\bar{4},\bar{4})$,
\end{itemize}

\begin{equation}
\Psi^c_{iL} = \begin{pmatrix}
U^{ic}_{r} & U^{ic}_{g} & U^{ic}_{b} & V^{ic} \\ 
D^{ic}_{r} & D^{ic}_{g}  & D^{ic}_{b} & E^{+i}\\
u^{ic}_{r} & u^{ic}_{g} & u^{ic}_{b} & \nu^{ic} \\ 
d^{ic}_{r} & d^{ic}_{g}  & d^{ic}_{b} & e^{+i} \\   
\end{pmatrix}_L \;\;\;,
\Psi^c_{iR} = \begin{pmatrix}
u^{ic}_{r} & u^{ic}_{g} & u^{ic}_{b} & \nu^{ic} \\ 
d^{ic}_{r} & d^{ic}_{g}  & d^{ic}_{b} & e^{+i} \\ 
U^{ic}_{r} & U^{ic}_{g} & U^{ic}_{b} & V^{ic} \\ 
D^{ic}_{r} & D^{ic}_{g}  & D^{ic}_{b} & E^{+i}\\  
\end{pmatrix}_R
\;, 
\label{antifermion}
\end{equation}\\
where $i=1,2,3$ are the generation index, so $\Psi_1, \Psi_2, \Psi_3$ are the fermionic fields of the first, second, and third generation respectively. Hence, $u^i=u,c,t$ are the quarks up, charm and top, $d^i=d,s,b$ are the quarks down, strange and bottom, $\nu^i=\nu_e, \nu_\mu, \nu_\tau$ are the electron neutrino, muon neutrino and tau neutrino, and $e^i=e, \mu, \tau$ are the electron, muon and tau lepton. The superscript $c$ denotes the corresponding charge conjugate state (antiparticle).    
Here we employed again a matrix notation in which $SU(4)_C$ indices run horizontally and $SU(4)_{RL}$ indices run vertically. The particle assignment shown in (\ref{fermion},\ref{antifermion}) is straightforwardly anomaly-free since left-handed and right-handed fermions are in the same representation of $SU(4)_C\times SU(4)_{RL}$  \cite{anomalies}.       

\begin{table*}[h!]
\scriptsize\centering
\label{fermreps}
\begin{tabular}{ccc}
\cline{1-3}
\multicolumn{1}{|c|}{$\left(SU(4)_C \times SU(4)_{RL}\right)$}                  & \multicolumn{1}{c|}{$\left(SU(4)_C \times SU(2)_L \times SU(2)_R\right)_{U(1)_\beta}$ } & \multicolumn{1}{c|}{ $\left(SU(3)_C \times SU(2)_L\right)_{U(1)_Y}$ }   \\ \cline{1-3}
\multicolumn{1}{|c|}{\multirow{2}{*}{Fermions (4,4)}} & \multicolumn{1}{c|}{ $(4,2,1)_{+1}$} & \multicolumn{1}{c|}{$(3,2)_{\frac{1}{3}} \; \oplus \;
(1,2)_{-1}$}  \\ \cline{2-3}
\multicolumn{1}{|c|}{}                  & \multicolumn{1}{c|}{$(4,1,2)_{-1}$} & \multicolumn{1}{c|}{$(3,1)_{\frac{4}{3}} \; \oplus \; (3,1)_{-\frac{2}{3}} 
 \; \oplus \; (1,1)_{-2} \; \oplus \; (1,1)_{0}$}  \\ \cline{1-3} 
 \multicolumn{1}{|c|}{\multirow{2}{*}{Anti-fermions ($\bar{4},\bar{4}$)}} & \multicolumn{1}{c|}{ $(\bar{4},2,1)_{-1}$} & \multicolumn{1}{c|}{($\bar{3},2)_{-\frac{1}{3}} \; \oplus \;
(1,2)_{1}$}  \\ \cline{2-3}
\multicolumn{1}{|c|}{}                  & \multicolumn{1}{c|}{$(\bar{4},1,2)_{+1}$} & \multicolumn{1}{c|}{$(\bar{3},1)_{-\frac{4}{3}} \; \oplus \; (\bar{3},1)_{\frac{2}{3}} \; \oplus \;
(1,1)_{2} \; \oplus \; (1,1)_{0}$} \\ \cline{1-3}
\end{tabular}
\caption{Fermionic particles and antiparticles representations at different energy scales.}
\end{table*} 

\section{Summary and discussion}

We can check that all fermions are transformed correctly under the SM gauge group at low energy scales, just as all known charges are well-assigned. Through the decomposition of $(4,4)$ and $(\bar{4},\bar{4})$ we find the representations of fermions at different energy scales (Table \textcolor{purple}{6}).   
Furthermore, the quantum numbers and charges of all fermions are displayed in Tables \textcolor{purple}{4} and \textcolor{purple}{5}.\\

Note that, in order to complete the multiplets (\ref{fermion},\ref{antifermion}), we introduce new particles which does not exist neither in the Standard Model nor in the Pati-Salam model. These exotic particles are heavy states which are transformed under the SM group at low energies of different way (Table \textcolor{purple}{6}). In particular, some exotic weakly interacting particles, which may have cosmological implications, are predicted. We have denoted them by $V$. They are transformed as $(1,1)_{0}$ (left-handed $V$) and $(1,2)_{-1}$ (right-handed $V$) under $G_{SM}$, and as $1_0$ under $G'_{SM}$. Notice as well that the value of the beta hypercharge distinguishes between the left-handed (right-handed) known fermions, with $\beta=+1$ ($-1$), and the left-handed (right-handed) exotic fermions, with $\beta=-1$ ($+1$).\\   

The matter and antimatter Lagrangian density of the Double SU(4) GUT Lagrangian can be written as
\begin{equation}
\mathcal{L}_M = \Tr\left[ \sum_{i=1}^3 \bar{\Psi}_{iL} \gamma^\mu D_\mu \Psi_{iL} + \bar{\Psi}_{iR} \gamma^\mu D_\mu \Psi_{iR} + \bar{\Psi}^c_{iL} \gamma^\mu D^c_\mu \Psi^c_{iL} + \bar{\Psi}^c_{iR} \gamma^\mu D^c_\mu \Psi^c_{iR} \right] \;,
\label{interaction}
\end{equation}
where the covariant derivative operators are
\begin{equation}
\gamma^\mu D_\mu \Psi= \gamma^\mu\left(\partial_\mu\Psi- i g_{4RL} A_{RL}\Psi -  i g_{4C}\Psi A_C \right) \;,
\end{equation}
\begin{equation}
\gamma^\mu D^c_\mu \Psi^c = \gamma^\mu\left(\partial_\mu\Psi^c+ i g_{4RL}A^T_{RL}\Psi^c +  i g_{4C}\Psi^c A^T_C \right)\;. 
\end{equation}\\

To sum up, in this work we have introduced the Double SU(4) GUT, a model based on the gauge group $SU(4)\times SU(4)\;\left(\times \mathcal{Z}_2\right)$. We have considered a pattern of symmetry breaking in which this group is broken via a Pati-Salam stage, and according to this pattern a complete set of generators have been constructed. 
Using the set of generators built, the gauge boson matrices have been calculated and we have defined convenient quantum numbers in order to label the states and to derive the extended Gell-Mann Okubo relation. Finally, the fermion content of the minimal model has also been studied.\\ 

On one hand, the unification through the gauge group $SU(4)\times SU(4)\times\mathcal{Z}_2$ allows us to unify all gauge coupling constants and to express the parity violation in the electroweak theory as a consequence of the spontaneous symmetry breaking. 
Furthermore, as shown in this work, in the framework of the Double SU(4) GUT all particles (antiparticles) of a generation are unified in two multiplets of a single representation $(4,4)$ ($\left(\bar{4},\bar{4}\right)$) of $SU(4)_C\times SU(4)_{RL}$. In order to complete these multiplets, exotic quarks and leptons are predicted. \\

On the other hand, the Double SU(4) model may have implications in Cosmology as is common in GUT. For instance, this model can produce significant baryogenesis depending on the particular symmetry breaking framework since the Sakharov condition of baryon number violation is increased by the double SU(4) interaction. In addition, the exotic WIMP right-handed $V$, and the super-WIMPs right-handed neutrino and exotic left-handed lepton $V$ are Dark Matter candidates \cite{Feng,Banerjee} which appear mandatorily in the minimal model.\\

More detailed investigation should be done in order to analyze the phenomenological consequences of this GUT model, and to realize the stages of symmetry breaking through for example, an extended Higgs mechanism. Further study of the Higgs sector of the model should also be performed to obtain the Yukawa terms of the Lagrangian and the particle mass spectrum. \\

\subsection*{Acknowledgments}
This work has been supported by the MINECO (Spain) projects FIS2014-52837-P, FIS2016-78859-P (AEI/FEDER, UE), and Consolider-Ingenio MULTIDARK CSD2009-00064, and the MECD Collaboration fellowship (Spain).

\newpage
\appendix
\section{Structure constants of SU(4)}\label{structurefuncappen}

The generators of $SU(4)_C$ and $SU(4)_{RL}$ constructed in section \ref{Groupconstruction} obey the characteristic commutation relations \ref{conmut} given by the structure constants $f_{abc}$ displayed in table \ref{tabsc}.
\begin{equation}
\comm{T_a}{T_b}=i\sum_j f_{abj} T_j \Longrightarrow 
f_{abc}= -2i \Trace(T_c\comm{T_a}{T_b})
\end{equation}
\begin{table*}[h!]
\small\centering
\begin{tabular}{|c|c|c|c|c|} \hline
a & b & c & $f^{C}_{abc}$ & $f^{RL}_{abc}$ \\\hline
1 & 2 & 3 & 1 & 1 \\\hline
1 & 4 & 7 & $\frac{1}{2}$ & $\frac{1}{2}$ \\\hline
1 & 5 & 6 & $-\frac{1}{2}$ & $-\frac{1}{2}$ \\\hline 
1  & 9 & 12 & $\frac{1}{2}$ & $\frac{1}{2}$ \\\hline
1  & 10 & 11 & $-\frac{1}{2}$ & $-\frac{1}{2}$ \\\hline
2  & 4 & 6 & $\frac{1}{2}$ & $\frac{1}{2}$ \\\hline
2  & 5 & 7 & $\frac{1}{2}$ & $\frac{1}{2}$ \\\hline
2  & 10 & 12 & $\frac{1}{2}$ & $\frac{1}{2}$ \\\hline
2  & 9 & 11 & $\frac{1}{2}$ & $\frac{1}{2}$ \\\hline
3  & 6 & 7 & $-\frac{1}{2}$ & $-\frac{1}{2}$ \\\hline
3  & 4 & 5 & $\frac{1}{2}$ & $\frac{1}{2}$ \\\hline
3  & 11 & 12 & $-\frac{1}{2}$ & $-\frac{1}{2}$ \\\hline
3  & 9 & 10 & $\frac{1}{2}$ & $\frac{1}{2}$ \\\hline
4  & 9 & 14 & $\frac{1}{2}$ & $\frac{1}{2}$ \\\hline
4  & 5 & 8 & $\frac{\sqrt{3}}{2}$ & $-\frac{1}{2}$ \\\hline
5  & 10 & 14 & $\frac{1}{2}$ & $\frac{1}{2}$ \\\hline
6  & 7 & 8 & $\frac{\sqrt{3}}{2}$ & $-\frac{1}{2}$ \\\hline
4  & 10 & 13 & $-\frac{1}{2}$ & $-\frac{1}{2}$ \\\hline
5  & 9 & 13 & $\frac{1}{2}$ & $\frac{1}{2}$ \\\hline
7  & 11 & 13 & $\frac{1}{2}$ & $\frac{1}{2}$ \\\hline
7  & 12 & 14 & $\frac{1}{2}$& $\frac{1}{2}$ \\\hline
6  & 11 & 14 & $\frac{1}{2}$& $\frac{1}{2}$ \\\hline
6  & 12 & 13 & $-\frac{1}{2}$& $-\frac{1}{2}$ \\\hline
9  & 10 & 15 & $\frac{\sqrt{6}}{3}$ & $\frac{\sqrt{2}}{2}$ \\\hline
8  & 13 & 14 & $-\frac{\sqrt{3}}{3}$ & 1 \\\hline
8  & 11 & 12 & $\frac{\sqrt{3}}{6}$ & $\frac{1}{2}$ \\\hline
8  & 9 & 10 & $\frac{\sqrt{3}}{6}$ & $\frac{1}{2}$ \\\hline
13  & 14 & 15 & $\frac{\sqrt{6}}{3}$ & 0 \\\hline
11  & 12 & 15 & $\frac{\sqrt{6}}{3}$ & $\frac{\sqrt{2}}{2}$ \\\hline
4  & 5 & 15 & 0 & $\frac{\sqrt{2}}{2}$ \\\hline
6  & 7 & 15 & 0 & $\frac{\sqrt{2}}{2}$ \\\hline
\end{tabular}\label{tabsc}
\caption{Non-zero independent totally antisymmetric structure constants $f_{abc}$ of $SU(4)_C$ $\left(f_{abc}^C\right)$ and $SU(4)_{RL}$ $\left(f_{abc}^{RL}\right)$. Note that only the structure constants which involve 8 or 15 indices satisfy $f^{C}_{abc}\neq f^{RL}_{abc}$.}
\end{table*}

\newpage
\bibliographystyle{unsrt}

\end{document}